\documentclass[twocolumn,amssymb,footinbib, aps, prb,reprint]{revtex4-1}

\usepackage{todonotes}
\usepackage{bm}
\usepackage{graphicx}
\usepackage{amsthm,amssymb}
\usepackage{amsmath}
\usepackage{datetime}

\usepackage{hyperref}
\hypersetup{colorlinks,
	urlcolor=[rgb]{0,0.6,0.6},
	linkcolor=[rgb]{0,0.0,0.8}}

\newcommand{\vf}{v_\text{\tiny $F$}}

\graphicspath{{figures/}}

\begin{document}

	\title{Robustness of ballistic transport in antidot superlattices}

	\author{George Datseris}
	\email{george.datseris@ds.mpg.de}
	\affiliation{Max Planck Institute for Dynamics and Self-Organization}
	\author{Theo Geisel}
	\affiliation{Max Planck Institute for Dynamics and Self-Organization}
	\author{Ragnar Fleischmann}
	\affiliation{Max Planck Institute for Dynamics and Self-Organization}
	\date{\today}

	\begin{abstract}

		The magneto-resistance of antidot lattices shows pronounced peaks, which became a hallmark of ballistic electron transport. While most studies agree that they reflect the interplay of regular and chaotic motion in the quasi-classical dynamics, the exact mechanism has been surprisingly controversial. Inspired by recent experiments on graphene antidot lattices showing that the effect survives strong impurity scattering, we give a new explanation of the peaks linked to a fundamental relation between collision times and accessible phase space volumes, accounting for their robustness. Due to the fundamental nature of the mechanism described it will be relevant in many mesoscopic transport phenomena.

	\end{abstract}

	\maketitle

	Antidot superlattices are nanostructured artificial crystals of repellers in high mobility two-dimensional electron gases (2DEGs).
	They were first fabricated by ion-beam implantation~\cite{Ensslin1990} or by etching periodic arrays of holes with periods of a few hundred nanometers into the 2DEG of AlGaAs/GaAs heterostructures~\cite{Weiss1991} and by modulation of the 2DEG by nanostructured lateral metal gates~\cite{Lorke1991}.
	Since then antidots were realized in many different  materials~\cite{Xiao2002, Wordenweber2004, Neusser2008}, and recently also in graphene~\cite{Sandner2015, Yagi2015, Oka2019} and topological insulators~\cite{Maier2017}.
	They are a prime example of devices showing features of ballistic transport: when the typical length scales of a device become smaller than the mean free path of the electrons, transport is no longer dominated by diffusion due to impurity, phonon or electron-electron scattering. Instead, transport is mainly affected by external forces or the potential of the device superstructure.

\begin{figure}[t!]
		\includegraphics[width=\columnwidth]{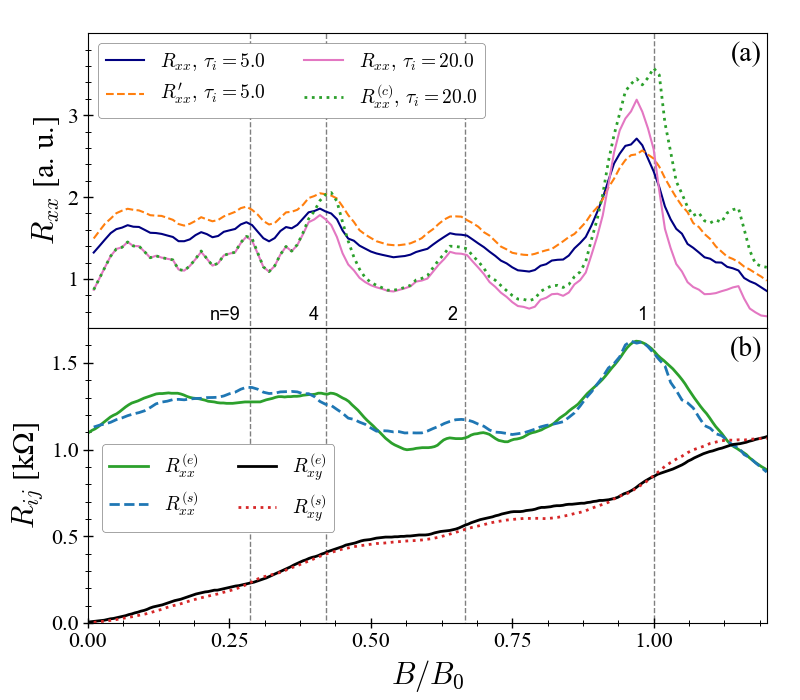}
		\caption{(a): Magnetoresistivity simulations of antidot superlattices (ADSLs) demonstrating commensurability peaks ($c=0.1, d_0=0.3, \tau_i = 5,20$). $R_{xx}'$ denotes simulations with boundary roughness with parameter $\varepsilon = 0.1$ (see Fig.~\ref{fig:orbit}b and the supplement for mathematical details). $R_{xx}^{(c)}$ is the resistivity of the chaotic part of the phase space. Vertical dashed lines indicate the commensurable fields $B_n$ (cf. Fig.~\ref{fig:orbit}). (b) Comparison of recent experiments of magneto- and Hall resistance $R^{(e)}$ by Sandner \textit{et al.}~\cite{Sandner2015} with simulations $R^{(s)}$ using parameters $c=0.2$, $d_0 = 0.3, \tau_i  = 2.5$ and $B_0 = 3.7\,$T (the prefactor was determined by a least squares fit).}
		\label{fig:magneto}
	\end{figure}

\begin{figure}[t!]
		\includegraphics[width=\columnwidth]{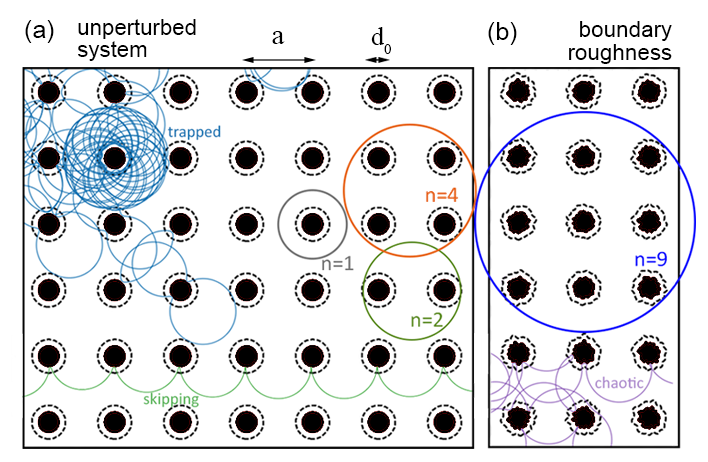}
		\caption{Contours of the potential $U$ (filled: $U \ge 1$, dashed: $U = 0$).  Antidots with boundary roughness are shown in (b), with parameter $\varepsilon = 0.1$ (see supplement). Examples of chaotic, skipping, as well as \emph{trapped} orbits at $B=1.0$ and pinned orbits enclosing $n = 1, 2, 4$ and $9$ antidots at the commensurable fields $B_n\in\{1,\, 0.66,\, 0.42,\, 0.285\}$) are also shown.}
		\label{fig:orbit}
	\end{figure}

	The most prominent of a number of ballistic transport effects observed in antidot lattices is their low temperature magneto-resistance at small magnetic fields which typically shows a series of pronounced peaks.
	It was realized~\cite{Weiss1991} that they occur at magnetic field values where collision-less circular cyclotron orbits encircling a certain numbers of antidots can exist in the superlattice (cf.~Figs.~\ref{fig:magneto} and \ref{fig:orbit}) and they are therefore known as \textit{commensurability peaks} (CPs).  At even lower temperatures, when the coherence length of the electrons grows, quantization of periodic orbits manifests itself in additional resistivity oscillations~\cite{Weiss1993,Silberbauer1994,Richter1995}.
	CPs in graphene have also been reproduced recently in tight-binding simulations of small antidot systems \cite{Power2017}.
	In the following we will consider the \textit{incoherent} ballistic transport, which can be analyzed in terms of quasi-classical dynamics~\cite{Fleischmann1992}.
	Such  quasi-classical resistivity peaks have even been observed at high magnetic fields in the fractional quantum hall regime, where compound quasi-particles, so called \textit{composite fermions}, experience a reduced effective magnetic field~\cite{Kang1995,Fleischmann1996}.

	While most studies agree that the CPs reflect the interplay of regular and chaotic quasi-classical dynamics, their origin has remained controversial~\cite{Fleischmann1992,Baskin1992,Schuster1994, Fliesser96, Ishizaka1997}.
	Recent experiments by Sandner \textit{et al.}~\cite{Sandner2015} on antidot lattices in monolayer graphene can give us new insight, as they reveal the existence of CPs despite strong impurity scattering. In the present letter we analyze the quasi-classical electron dynamics in graphene antidot lattices to interpret these magnetoresistance experiments and provide a new explanation for the mechanisms giving rise to CPs in general.

	The first mechanism proposed to explain CPs was based on the assumption that cyclotron orbits encircling antidots get \textit{pinned} by the antidots and do not contribute to conduction~\cite{Weiss1991}.
	It was shown theoretically that nonlinear resonances in soft wall models can give rise to such a pinning mechanism~\cite{Fleischmann1992}. Furthermore it was demonstrated that the resistivity contributions of the chaotic (i.e.~non pinned) trajectories alone can already produce CPs and that pinning is not the dominant effect~\cite{Fleischmann1992}.
	How the chaotic dynamics leads to CPs, however, was strongly disputed (see e.g.\ Refs.~\onlinecite{Fleischmann1992,Baskin1992,Schuster1994,Fliesser96, Ishizaka1997}).
	In any case one would assume that the mean impurity scattering time $\tau_i$ due to residual disorder in the substrate must be sufficiently long in order to observe features of chaotic ballistic dynamics and nonlinear resonances in transport measurements.
	More specifically it should fulfill $\omega \tau_i \gg 2\pi$, where $\omega$ is the typical frequency of the nonlinear resonances  (which in the antidot system is close to the cyclotron frequency $\omega_c$).
	In the graphene experiments by Sandner \textit{et al.}~\cite{Sandner2015} and Yagi \textit{et al.}~\cite{Yagi2015}, however, the scattering time is quite short and thus $\omega \tau_i$ only barely reaches $2\pi$ (see below for numeric values).

	Why do the experiments still exhibit CPs nevertheless? In this article we will show that the resistance peaks survive the impact of these short impurity scattering times due to a deep rooted connection of different dynamical properties. The nonlinear resonances, by their mere existence and simply because they are taking away a part of the chaotic phase space volume, are reducing the fastest chaotic timescale, i.e.~the mean time between (successive) collisions with antidots. A fact that we will proof using \textit{Kac's lemma}.  In contrast to the time scales of the nonlinear resonances, the mean collision time in the experiment is shorter or at least comparable to the impurity scattering time, and we argue that its reduction is what is observed in experiment as a resistance peak.
The discovered mechanism is very general and will therefore be applicable to a wide range of billiard-like mesoscopic systems: generically the intrinsic scattering time of the chaotic orbits is linked to nonlinear resonances simply by their phase space volume.

The paper is structured as follows. In section~\ref{sec:model} we first introduce the model we use to describe transport in graphene antidot lattices. Our model uses smooth antidot potentials and includes bulk and boundary disorder. We show that it is capable of quantitatively reproducing the experimental findings very well. In section~\ref{sec:analysis} we analyse the numerical simulations and find that the resistance peaks correspond to minima in the fastest dynamical time scale, the mean collision time. To conclude our analysis and to study this fast timescales in detail,  we switch to a billiard model with hard walled antidots.  The billiard does not as closely reproduce the experimental findings but allows us to exhibit the underlying dynamical principles much clearer in a well defined and partially analytically treatable manner.
We end our discussion by showing in section~\ref{sec:stochastic} that even in a purely stochastic model of uncorrelated collisions with diffusely scattering antidots the minima in the collision time lead to resistivity peaks. This clearly demonstrates the robustness of the effect.

	\section{Model}\label{sec:model}

	We study the electron transport in the single particle picture using (quasi-)classical Hamiltonian dynamics~\cite{Richter1995}. We therefore model the carriers by a quasi-classical micro-canonical ensemble at the Fermi energy $E_F$. The Hamiltonian is $\mathcal{H} = \mathcal{K}(p_x, p_y) + \mathcal{U}(x,y)$, where $\mathcal{K} = v_F\sqrt{p_x^2 + p_y^2}$ follows from the Dirac approximation of the dynamics of electrons in graphene with Fermi velocity $v_F$~\cite{CastroNeto2009}. Our main conclusions apply equally well to the more common quadratic dispersion relation $\mathcal{K}\propto \mathbf{p}^2$, as will become apparent later.
	The magnetic field perpendicular to the graphene layer is introduced by \emph{minimal coupling}, $\mathbf{p} \to \mathbf{p} - q\mathbf{A}$, choosing the symmetric gauge $\mathbf{A}= -\tfrac{1}{2}\mathbf{r\times B} = \tfrac{1}{2}(-yB, xB, 0)$.
	We model the antidot super-lattice by a square array of localized, isotropic potential peaks of the form
	\begin{equation}
	U_a = \left(\tfrac{E_F}{c^4}\right) \left[\tfrac{d_0}{2} + c - r_a\right]^4
	\end{equation}
	if $r_a = \sqrt{(x-x_a)^2 + (y-y_a)^2} < \frac{d_0}{2} + c$ and 0 otherwise.
	$(x_a,y_a) = (na, ma)\, $ with $n, m \in \mathbb{N}$ is the center of the (nearest) antidot and $a$ is the super-lattice constant. $d_0$ is the antidot diameter (in the experiment by Sandner \textit{et al.}~\cite{Sandner2015} reported values are $d_0 \approx 25-30$nm with antidot spacing $a \approx 100$nm). The parameter $c$ defines a cut-off distance of the potential and it is also the parameter that controls the steepness of the antidots. The full potential is then the (infinite) sum of the above localized potentials over all antidot centers $\mathcal{U}= \sum_{n, m} U_a$.
	A contour plot of $\mathcal{U}$ is given in Fig.~\ref{fig:orbit}a. We want to stress that even though of course the diameter and smoothness of the antidots are important parameters, different functional forms (e.g. $\mathcal{U} \propto [\cos(2\pi x/a)\cos(2\pi y/a)]^\beta$, like in \cite{Fleischmann1992}) for the potential have very little impact on the resistivities, and no bearing at all on our main conclusions.

	We can add boundary roughness (which aims to model the antidot fabrication defects) to the above potential by making the antidot diameter depend pseudo-randomly on the position. Specifically, let $\tilde{x} = x - x_a$, $\tilde{y} = y - y_a$ and define $\phi = \arctan(\tilde{y}/\tilde{x})$. We can then write
	\begin{equation}
	d(\phi) = d_0 + d_0\varepsilon\sum_{\zeta = 1}^{M} \nu(x_a, y_a, \zeta)\times\sin(\zeta\phi)
	\end{equation}
	with $\nu$ being random numbers uniformly distributed in $[-0.5, 0.5)$ which are different for each combination of $x_a, y_a, \zeta$, so that on average $\langle d(\phi) \rangle = d_0$. $M$ is the total number of sine modes used (measure of the edge complexity) and $\varepsilon$ measures the relative boundary roughness. \footnote{In the numerical computations we used an array of $100\times M$ random numbers repeated periodically and used $M=16$ for all simulations shown in the manuscript.} A realization is shown in Fig.~\ref{fig:orbit}b. (Note that this boundary roughness is distinct from the impurity scattering, which we also included in our model, see eq.~\eqref{kubo1}.)

	Introducing dimensionless variables $x_i \to x_i/a$ (where $a$ is the antidot lattice spacing),  $v_i \to v_i/\vf$  and scaling the energy by the Fermi energy $E_F = \hbar \vf \sqrt{\pi n_e}$~\cite{CastroNeto2009} ($\approx 0.1-1$ eV in \cite{Sandner2015}), the Hamiltonian of our model becomes
	\begin{align}
	\mathcal{H} = \sqrt{\left(p_x + yB \right)^2 +
		\left(p_y - xB \right)^2  } + \mathcal{U}(x,y).
	\label{hamiltonian}
	\end{align}
	In the following we study the dynamics on the manifold $H=1$ (i.e. the Fermi energy). We scale the magnetic field by its value at the principal ($n = 1$) commensurability $B_0 = 2\hbar\sqrt{\pi n_e}/(ea)$, corresponding to a cyclotron diameter equal to the antidot lattice constant $a$. All times are given in units of $t_0=a/\vf$ and the cyclotron frequency is $\omega_c=2B$.

	Eq.~\eqref{hamiltonian} leads to the equations of motion $\dot{\mathbf{x}} = \partial \mathcal{H} / \partial\mathbf{p}, \dot{\mathbf{p}} = -\partial\mathcal{H}/\partial\mathbf{x}$. We then obtain equations for velocities instead of momenta,
	\begin{align}
	\dot{x} &= v_x \label{veleom_start}\\
	\dot{y} &= v_y \\
	\dot{v}_x & = v_y \left(\frac{v_x \frac{\partial\mathcal{U}}{\partial y} - v_y\frac{\partial \mathcal{U}}{\partial x} + 2B}{1-\mathcal{U}}\right) \\
	\dot{v}_y & = - v_x \left(\frac{v_x \frac{\partial \mathcal{U}}{\partial y} - v_y\frac{\partial \mathcal{U}}{\partial x} + 2B}{1-\mathcal{U}}\right) .
	\label{veleom_end}
	\end{align}
	Eqs.~\eqref{veleom_start}-\eqref{veleom_end} describe (classical) hyper-relativistic particles in magnetic field $B$ and potential $\mathcal{U}$. The velocity timeseries are obtained by numerical integration of the equations of motion using standard Runge-Kutta schemes.

	After obtaining the velocity time series we then calculate the conductivities and resistivities using the Kubo formalism~\cite{Kubo1957}. Following~\cite{Fleischmann1992} we write
	\begin{align}
	&\sigma_{ij} \sim \int_0^\infty e^{-t/\tau_i}   \langle v_i(t)v_j(0) \rangle_{E_F} \,dt \, ,
	\label{kubo1} \\
	&R_{ij} = \frac{\sigma_{ij}}{\sigma^2_{xx} + \sigma^2_{xy}}
	\label{resistivities}
	\end{align}
	with the conductivities $\sigma_{ij}$  ($ij = xx$ or $xy$) and the magneto- and Hall-resistivity $R_{xx}$ and $R_{xy}$, respectively (modulo some geometry prefactor).
	$ C_{ij}(t) \equiv \langle v_i(t)v_j(0) \rangle_{E_F}$ is the velocity correlation function (VCF), averaged over the available phase space at the Fermi energy. Impurity scattering is introduced in Eq.~\ref{kubo1} by assuming that the electron velocities are decorrelated by random scattering events which follow a Poisson distribution with mean time $\tau_i$~\cite{Fleischmann1992}. In the experimental paper~\cite{Sandner2015}, the authors, using a simple model, estimated  $\tau_i \approx 3.5$, however the best fit to our more elaborated numerical model yields $2.5\pm0.25$ (see below). We will therefore assume the latter to be the correct value. As $\omega_c\tau_i<2\pi$, we cannot assume that the regular orbits are actually pinned (with $\sigma_{ij}^{(r)}=0$ as in Ref.~\onlinecite{Fleischmann1992}), since they get scattered before they close. We therefore treated them within the Kubo formalism. They lead to Drude like contributions. By assuming $v_x = \cos(2Bt)$ and $v_y=\sin(2Bt)$ (i.e. regular cyclotron motion) the velocity correlation functions for the regular phase space are $ C_{xx}^{(r)}(t) = \cos(2Bt)/2$ and $C_{xy}^{(r)}(t) = \sin(2Bt)/2$.

	\section{Analysis}\label{sec:analysis}

	In the following discussion it is advantageous to divide the VCFs into contributions of the different regions of the mixed phase space.
	The chaotic part of the phase space has portion $g_c$ and its correlations $C_{ij}^{(c)}$ are decaying.
	On the other hand, the regular phase space with portion $g_r$ (where $g_c + g_r = 1$) has non-decaying correlations $C_{ij}^{(r)}$.
	We thus write
	$C_{ij} = (1-g_c)C_{ij}^{(r)} + g_cC_{ij}^{(c)}$, and subsequently
	\begin{align}
	\sigma_{ij} = (1-g_c)\sigma_{ij}^{(r)} + g_c\,\sigma_{ij}^{(c)}.
	\label{cond}
	\end{align}

	Note that the regular islands of pinned orbits (see Fig.~\ref{fig:orbit}) correspond to almost circular (quasi-)periodic orbits, which in a smooth antidot potential are stable against the application of an external electric field~\cite{Fleischmann1992}.
	Examples of these pinned orbits at the first four commensurabilities are shown in Fig.~\ref{fig:orbit}. Islands can also correspond to various forms of skipping~\cite{Baskin1992} orbits, an example is also shown in the figure. It has been argued that skipping orbits are essential for the creation of resistance peaks~\cite{Baskin1992,Schuster1994,Fliesser96,Ishizaka1997}.
	\emph{Boundary roughness}, which probably is present in the experiments of Ref.~\onlinecite{Sandner2015}, however, almost completely eliminates their existence without affecting the pinned orbits much.

	Figure~\ref{fig:magneto}a shows example magnetoresistivity curves for the unperturbed antidot lattice for various values of the impurity scattering time $\tau_i$. An excellent fit to the experiment in magneto- and Hall resistivity is achieved for a small scattering time $\tau_i=2.5$, as shown in Fig.~\ref{fig:magneto}b.
	Notice that the 4-peak is observable in both experiments and simulations, with characteristic ballistic time scale $T_R\approx\pi/0.44 \approx 7.1$ (in our units). This means that this ballistic feature is observable even when the impurity time is almost three times shorter than the associated ballistic time scale!

	The magnetoresistivity of an antidot lattice with boundary roughness ($R'_{xx}$), shown in Fig.~\ref{fig:magneto}a exhibits surprisingly little differences from the unperturbed lattice. This suggests that the skipping orbits have little impact on the CPs.
	Furthermore, the resistivity curve $R_{xx}^{(c)}$ of only the chaotic part of phase space, already fully reveals all CPs.
	It thus suffices to understand how the chaotic correlations change with the magnetic field to produce CPs.

	Chaotic orbits get \textit{trapped} near regular islands~\cite{Meiss1985,Geisel1987, Geisel1988, Tabor, Arnold1989, Fleischmann1992}, i.e.\ they follow the quasiperiodic motion of the regular orbits for long times, as illustrated in Fig.~\ref{fig:orbit}a.
	This leads to long-time tails in the correlations of the chaotic orbits.
	It was therefore argued that they would give rise to valleys rather than peaks in the magnetoresistivity~\cite{Fliesser96}.
	We will show, however, that this effect is overcompensated by an initially accelerated correlation decay as a direct consequence of phase space volume conservation.

\begin{figure}[t!]
		\includegraphics[width=\columnwidth]{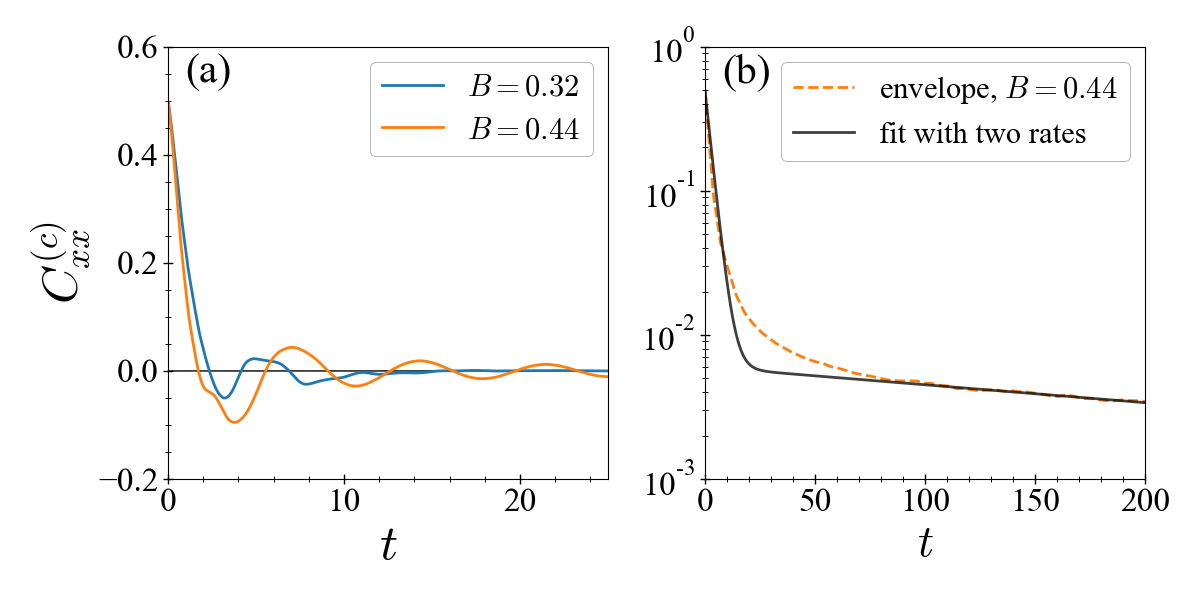}
		\caption{(a): Velocity correlation function of chaotic orbits for $B=0.32, 0.44$ and $ c=0.1,d_0=0.3$. (b): Envelope of $C^{(c)}_{xx}$ and a plot of $p \exp(-t/\tau_\text{fast}) + (1-p)\exp(-t/\tau_\text{slow})$ with $p=0.985,\tau_\text{fast}=3.0,\tau_\text{slow}=350.0$. While for non-commensurate magnetic fields ($B=0.32$) there is a fast exponential decay, for commensurate magnetic fields ($B= 0.44$) there is an additional slow decay as seen in the envelope.}
		\label{fig:cors}
	\end{figure}

	To do so, let us first inspect examples of the chaotic VCFs $C_{ij}^{(c)}(t)$  in Fig.~\ref{fig:cors}.
	In the absence of regular islands in phase space ($B=0.32$) we observe a single, fast correlation decay. In the presence of islands ($B=0.44$),  we see long time tails in the correlations, originating from trapping.
	From fig.~\ref{fig:cors}b, showing the envelope of the autocorrelation, however, it becomes clear, that the decay is an overlap of a fast and a slow component. We may think of it in the form
	\begin{equation}
	C_{xx}^{(c)}(t) \approx g_\text{\tiny col} C_\text{\tiny fast}(t) + g_\text{\tiny trap} C_\text{\tiny slow}(t)
	\label{envelope}
	\end{equation}
	(and similarly for the cross-correlation $C_{xy}$) where $g_\text{\tiny trap}$ is an appropriate measure of the trapping regions in the chaotic sea, and  $g_\text{\tiny col}=g_c-g_\text{\tiny trap}$ is its complement, i.e.\ the phase space region with strong chaotic scattering at the antidots (``\textit{collisions}'').
	The black line of Fig.~\ref{fig:cors}b illustrates this division by assuming exponential correlation decays with a fast and a slow time constant, $\tau_\text{\tiny fast} = 3$ and $\tau_\text{\tiny slow} = 350$, with $g_\text{\tiny trap} = 1.5\%$.
	At this point, the detailed values are of no concern.
	What we see is that $\tau_\text{\tiny slow}$ is orders of magnitude larger than $\tau_i$.
	The slow decay rate is therefore almost zero in comparison and the portion of trapped chaotic orbits is just a (small) correction to the portion of pinned orbits.
	The main contribution to the CPs stems from the change in the fast decay, which reflects the dynamics of the highly chaotic part of the phase space, a finding which is also reported by recent quantum simulations on graphene antidots by Power \textit{et al.}~\cite{Power2017}.
	This is also in accordance with the magnetic focusing mechanism between successive collisions with an antidot and its neighbours that has been argued to increase the diffusivity of chaotic orbits in Ref.~\onlinecite{Ando1999}.
	The fast correlation decay can be studied best in a billiard model with infinitely steep antidot walls, like in the Sinai billiard~\cite{Sinai1970}, which allows for an analytical treatment and should be a good approximation for these highly chaotic trajectories.

	In the remainder of this letter we will therefore study the periodic \textit{Sinai billiard} (PSB), also known as \emph{periodic Lorentz gas}, which is probably the most prominent example of a low-dimensional ergodic system \cite{Sinai1970, Weiss1991, Fliesser1996}.
	The PSB is the infinite steepness limit of the antidot super-lattice: the particles perform true free flight and are reflected specularly when colliding with the disks (i.e. the antidots).
	We simulate the PSB using an open source software we developed~\cite{Datseris2017}.
	We denote the collision times in the PSB by $t_\kappa$ and the mean collision time by $\kappa$, having in mind that the time scale of the fast correlation decay $\tau_\text{\tiny fast}$ in the original model is approximated by $\kappa$.

\begin{figure}[t!]
		\includegraphics[width=\columnwidth]{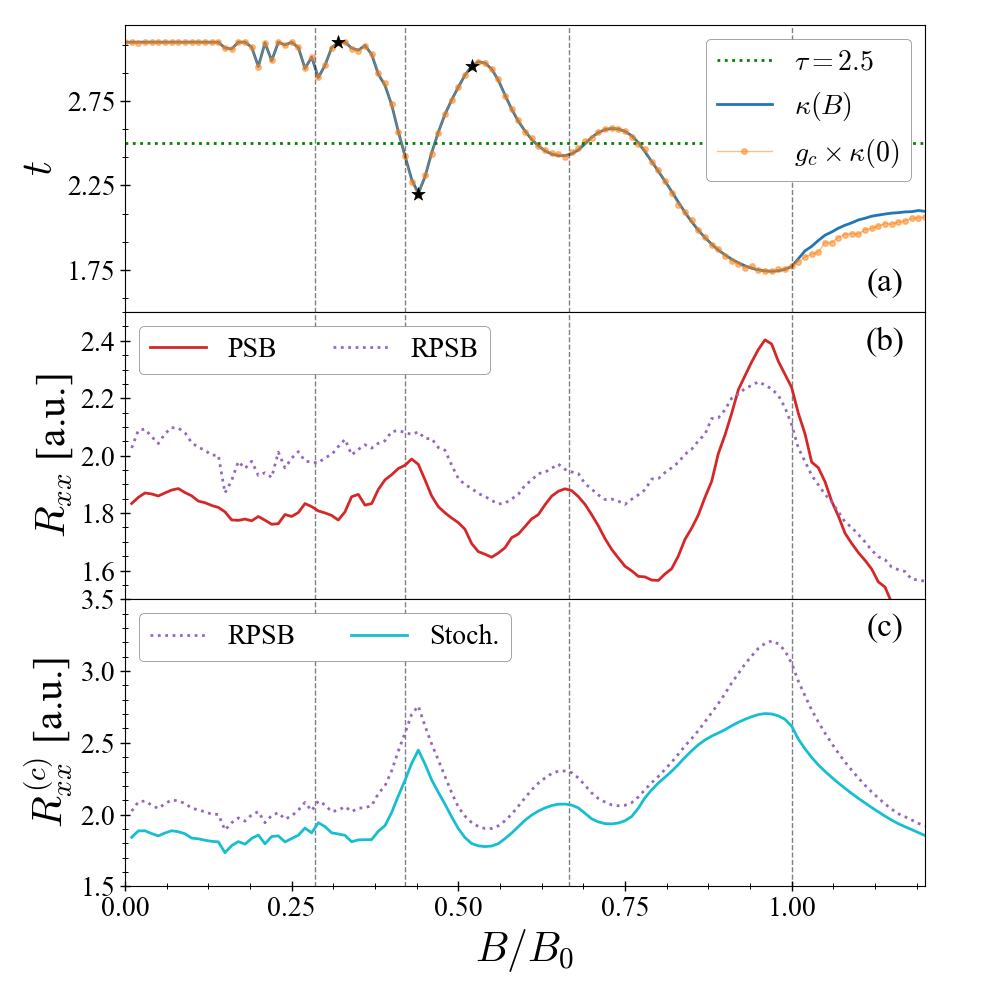}
		\caption{(a): Average collision time $\kappa$ in the periodic Sinai billiard (PSB) versus $B$. It coincides with the portion of chaotic orbits $g_c$ times $\kappa(0)$. (b) The resistivity curve of the random PSB (RPSB) does not differ much from that of the PSB. (c) The RPSB resistivity curve is approximated sufficiently well by the stochastic model (labeled ``Stoc.''). All curves are calculated with $\tau_i = 2.5, d_0 = 0.3$.}
		\label{fig:psb}
	\end{figure}

	Figure~\ref{fig:psb}a shows $\kappa$ as a function of the magnetic field. Remarkably, it exhibits pronounced valleys at the magnetic field values of the CPs in $R_{xx}$ as shown in Fig.~\ref{fig:psb}b for the billiard model.
	In the whole $B$ range $\kappa$ is comparable with $\tau_i$ and even smaller at the commensurable fields.
	The valleys in $\kappa(B)$ are the origin of the CPs. It is intuitive that more frequent collisions lead to reduced conductivity and thus increased $R_{xx}$, and we will further confirm this argument below using a simplified stochastic model.
	But first we show that the structure of $\kappa(B)$ has a deep connection with the mixed nature of the phase space (i.e. the coexistence of regular and chaotic phase space regions):
	Fig.~\ref{fig:psb}a indeed shows that it is directly proportional to the portion of the chaotic part of phase space  $g_c(B)$.

	We can understand this striking fact using \textit{Kac's lemma} \cite{Kac1947, MacKay1994, Meiss1997, Altmann2008}, which is a direct consequence of phase space volume conservation in Hamiltonian systems. We obtain a map of the flow $\Phi^t$ by discretizing in time, $f := \Phi^{\Delta t_\varepsilon}$, with $\Delta t_\varepsilon \sim \varepsilon$. Kac's lemma states that the mean number of iterations $\langle N_\mathcal{S} \rangle$ needed to return to a compact subset $\mathcal{S}$ of the phase space $\mathcal{M}$ is given by
	\begin{equation}
	\langle N_\mathcal{S}\rangle(B) = \frac{\mu(\mathcal{M}_{acc};B)}{\mu(S;B)},
	\label{recurrence}
	\end{equation}
	where $\mu(\cdot)$ is the phase space measure and $\mathcal{M}_{acc}$ the part of the phase space accessible to orbits starting in $\mathcal{S}$. Let $\mathcal{T}_\mathcal{S} (B) = \Delta t_\varepsilon\times  \langle N_\mathcal{S} \rangle(B)$ denote physical time instead of map iterations.
	Let $\mathcal{W}$ be a circle of radius $d_0/2 + \varepsilon$ concentric to the antidot and define $\mathcal{S}_\varepsilon \subset \mathcal{M}$ such that
	\begin{equation}
	\mathcal{S}_\varepsilon = \{\vec{x}, \vec{v}: \vec{x} \in \mathcal{W} \; \text{and} \; \vec{v} \cdot \vec{\eta}(\vec{x}) < 0 \}
	\end{equation}
	where $\vec{\eta}(\vec{x})$ is the vector normal to $\mathcal{W}$. The mean collision time $\kappa$ of the PBS is exactly $\mathcal{T}_\mathcal{S}$ in the limit $\varepsilon \to 0$.

	To find $\mu(\mathcal{S};B)$ we first realize that it does not depend on the magnetic field, $\mu(\mathcal{S};B) = \mu(\mathcal{S};0) = \mu(\mathcal{S})$.
	This is due to the infinitesimal width of $\mathcal{S}$, over which motion can always be approximated by a straight line for all finite magnetic fields values (i.e. equalling the  magnetic field free case).
	Then, using \eqref{recurrence} at $B=0$, we have
	\begin{equation}
	\mathcal{T}_\mathcal{S}(0) =\left( \frac{\Delta t_\varepsilon}{\mu(\mathcal{S_\varepsilon})} \right)  \mu(\mathcal{M}_{acc};0) = \left( \frac{\Delta t_\varepsilon}{\mu(\mathcal{S_\varepsilon})} \right),
	\label{convergence}
	\end{equation}
	because the PSB without magnetic field is fully ergodic and thus $\mu(\mathcal{M}_{acc};0)=\mu(\mathcal{M})=1$.
	By substitution we get
	$
	\mathcal{T}_\mathcal{S}(B) = \mu(\mathcal{M}_{acc};B)\times \mathcal{T}_\mathcal{S}(0)
	$.
	For small enough magnetic fields all chaotic orbits (and up to measure 0 only those) collide with the antidots.
	In the limit $\varepsilon\to 0$ both $\Delta t_\varepsilon$ and $\mu(\mathcal{S}_\varepsilon) $ go to 0 linearly with $\varepsilon$, therefore their ratio converges, i.e. $\mathcal{T}_\mathcal{S} \to \kappa$.
	Since by definition $\mu(\mathcal{M}_{acc})=g_c$, we find that the mean collision time is given by the fraction of chaotic orbits as a function of the magnetic field $B$, times the mean collision time at $B=0$
	\begin{equation}
	\kappa(B) = g_c(B)\times  \kappa(0).
	\end{equation}

	From this derivation it becomes also clear that the collision times are not sensitive to boundary roughness. This is intuitive, since the dynamics is already highly chaotic. To confirm this we also simulated a ``rough'' Sinai billiard where the particle gets reflected in a random angle at collision with the antidot (RPSB), corresponding to strong boundary roughness. The changes in the magnetoresistivity are only small indeed as shown in Fig.~\ref{fig:psb}b.

	\subsection{Stochastic Model}\label{sec:stochastic}

	\begin{figure}[t]
		\includegraphics[width=\columnwidth]{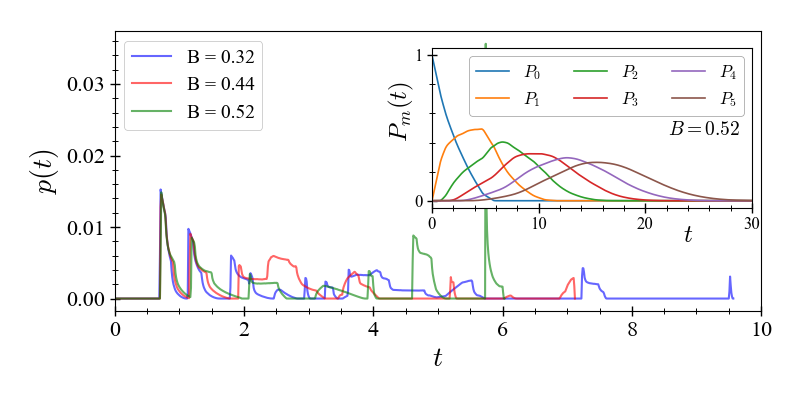}
		\caption{Probability density function of the collision times $p(t_\kappa)$ in the rough periodic Sinai billiard for the magnetic fields values marked by black stars in Fig.~4 of the manuscript, i.e.\ $B\in\{0.32, 0.44, 0.52\}$.  The inset shows numerically computed probabilities $P_m(t;B)$ for $B=0.52$.}
		\label{fig:hist}
	\end{figure}

	For completeness we now conclude our argumentation that the CPs arise due to the collision times by linking the collision times to the resistivities in a simplified purely stochastic analytic model for the motion in the billiard with boundary roughness (RPSB).
	We approximate the dynamics by a renewal process of stochastic scattering events (the distribution of the scattering times reflects the geometry of the antidot lattice) and free cyclotron motion in between scattering events. Each scattering event with a rough boundary leads to a random change in the velocity angle
	\begin{equation}
	\alpha=\pi +2 \psi,
	\end{equation}
	where $\psi$ is the random angle formed by the normal of the boundary segment and the velocity vector of the trajectory undergoing the scattering event. Under the assumption that all spatial orientations of the boundary segments are equally probable the probability density of hitting a segment with angle $\psi$ is
	\begin{equation}
	h(\psi)=\tfrac{1}{2} \cos \psi ,\,\, \mathrm{with}\,\, \psi\in \left[ -\pi/2,\pi/2 \right].
	\label{eq:Ppsi}
	\end{equation}
	Notice that this stochastic model is not the same as the RPSB; in the latter there is a strong correlation between the reflected angle and the subsequent collision time. This correlation does not exist in the stochastic model, as well as any knowledge of phase space volumes.

	The velocity correlation functions are
	\begin{align}\label{eq:corr1}
	S_{xx}&=\left< v_y(t) v_x(0) \right>=v_0^2\left< \cos(\varphi(t))  \cos(\varphi(0)) \right>\\
	S_{yx}&=\left< v_y(t) v_x(0) \right>=v_0^2\left< \sin(\varphi(t))  \cos(\varphi(0)) \right>,
	\end{align}
	where $v_0$ is the constant absolute value of the velocity ($v_0=1$ in our units) and the ensemble average $\left<\cdot\right>$  reduces to an average over the initial angle $\varphi_0$ of the velocities and the collision events. In the free propagation in-between events the velocity angle changes by $\Delta\varphi(t) = \omega\Delta t$ with $\omega=2B$ the cyclotron frequency.
	The contribution to the correlation functions of all trajectories that have scattered (exactly) $m$ times up to time $t$ is given by
	\begin{widetext}
		\begin{align}\label{eq:cxxbr}
		S_{xx}(t|m)=&P_m(t;B) \int_{-\pi}^{\pi} \mathrm{d} \varphi_0  \int_{-\pi/2}^{\pi/2} \mathrm{d} \psi_1 \dots \int_{-\pi/2}^{\pi/2} \mathrm{d} \psi_m \dfrac{\cos\varphi_0}{2\pi} \cos\left(\omega t +\varphi_0+m \pi +2 \sum_{i=1}^m \psi_i\right) \dfrac{\prod_{i=1}^{m}\cos \psi_i}{2^m}\\
		S_{yx}(t|m)=&P_m(t;B) \int_{-\pi}^{\pi} \mathrm{d} \varphi_0  \int_{-\pi/2}^{\pi/2} \mathrm{d} \psi_1 \dots \int_{-\pi/2}^{\pi/2} \mathrm{d} \psi_m \dfrac{\cos\varphi_0}{2\pi} \sin\left(\omega t +\varphi_0+m \pi +2 \sum_{i=1}^m \psi_i\right) \dfrac{\prod_{i=1}^{m}\cos \psi_i}{2^m},
		\end{align}
	\end{widetext}
	where $P_m(t;B)$ is the probability that a trajectory has scattered exactly $m$ times up to time $t$.
	The $\psi_i$ integrals can be easily carried out successively. Using the notation $\phi_m=\omega t +\varphi_0+m \pi +2 \sum_{i=1}^m \psi_i$, we write \begin{align*}\cos(\phi_m)=&\cos(\phi_{m-1}+2\psi_m + \pi)\\=&-\cos (\phi_{m-1}) \cos (2\psi_m)+\sin (\phi_{m-1}) \sin (2\psi_m)\end{align*}
	and similarly
	\begin{align*}\sin(\phi_m)= -\cos \left(\phi _{m-1}\right) \sin (2\psi_m)-\sin \left(\phi _{m-1}\right) \cos (2\psi_m).
	\end{align*}
	With
	\begin{align*}&	\frac{1}{2}\int_{-\pi/2}^{\pi/2} \cos(x)\cos(2x) \mathrm{d}x=\frac{1}{3},\\& \frac{1}{2}\int_{-\pi/2}^{\pi/2} \cos(x)\sin(2x) \mathrm{d}x=0\end{align*} and $$\frac{1}{2 \pi}\int_{-\pi}^{\pi} \cos(\varphi_0) \cos(\omega t+\varphi_0) \mathrm{d} \varphi_0=\frac{1}{2} \cos(\omega t)$$ we find $C_{xx}(t|m)=\frac{1}{2}\left(-\frac{1}{3}\right)^m \cos(\omega t)$ and $C_{yx}(t|m)=\frac{1}{2}\left(-\frac{1}{3}\right)^m \sin(\omega t)$. This finally allows us to write the correlation functions as the sum of these contributions:
	\begin{align}
	S_{xx}(t)&=\dfrac{1}{2}\sum_{m=0}^{\infty} \left(-\frac{1}{3}\right)^m P_m(t;B) \cos(\omega t)\\
	S_{yx}(t)&=\dfrac{1}{2}\sum_{m=0}^{\infty} \left(-\frac{1}{3}\right)^m P_m(t;B) \sin(\omega t).
	\end{align}

	The probability density function (pdf)  of the collision time $p(t)$ is a complicated function that is very sensitive to changes in the magnetic field, as shown in Fig.~\ref{fig:hist}. Therefore it is hard to find a useful analytical model for the probability $P_m(t;B)$. Only in the limit of large $m$ they are well approximated by Gaussians (see e.g. Ref.~\onlinecite{Mitov2014}). We are, however, mainly interested in short times and thus small $m$. Therefore we choose to calculate the $P_m(t;B)$ numerically using the following method: let $q_m(t)$ denote the pdf for the $m$-th scattering event ($m\ge1$) occuring at time $t$.
	It can be calculated through recursive convolution with the pdf $p(t)$,
	$$q_m(t) = \int_0^tdz\; p(z) q_{m-1}(t-z)$$
	with $q_1(t) \equiv p(t)$.
	Then $P_m(t)$ can be found from $q_m$ using the survival function $W(t)=\int_t^\infty p(t') dt'$
	\begin{equation}
	P_m(t) = \int_0^t dz \; W(z) q_m(t-z)
	\end{equation}
	for $m>0$ and $P_0(t) \equiv W(t)$.
	Examples of $P_m(t)$ are shown in the inset of Fig.~\ref{fig:hist}.

	Our stochastic model reproduces the contribution of the chaotic orbits to the CPs in the RPBS in very good approximation, as shown in Fig.~\ref{fig:psb}c. This confirms our claim that the CPs are due merely to the distribution of collision times in the antidot lattice.

	\bigskip 
	\section{Conclusions}

	In conclusion, we have explained recent magnetotransport experiments on graphene antidot lattices using appropriate quasiclassical electron dynamics. (And we note that our approach can also be applied to the analysis of the Hall-resistivity which shows a quenched or even negative Hall-effect at very small magnetic fields). We found that ballistic transport features like the CPs are visible in the resistivities even though the mean free time due to impurity scattering is so short that it is comparable to the fastest timescales of the chaotic dynamics. We showed that this striking robustness of the commensurability features can be understood by the fact that the fast chaotic time scale, the collision time, is reduced by the mere existence of stable islands in a mixed phase space, which reduce the chaotic phase space volume (Kac's Lemma).
	By this we solved a decades old riddle on the influence of nonlinear resonances on magnetotransport in antidot superlattices. Finally, due to the fundamental nature of the mechanism linking the time scales, which only depends on the basic properties of chaotic Hamiltonian systems with mixed phase space, it will be  generally applicable to a wide range of mesoscopic systems.

	\begin{acknowledgments}
		We thank Andreas Sandner and Jonathan Eroms very much for granting us access to their experimental data and fruitful discussions. We thank Stephan Eule for helpful discussions.
	\end{acknowledgments}
	\bibliographystyle{apsrev4-1}
	\bibliography{paper}

\end{document}